\documentclass[conference]{IEEEtran}
\IEEEoverridecommandlockouts

\usepackage{mathrsfs}
\usepackage{mathrsfs}
\usepackage{mathrsfs}
\usepackage{amssymb}
\usepackage{amsthm}
\usepackage{amsmath}
\usepackage{amsfonts}
\usepackage{mathrsfs}
\usepackage{amssymb}
\usepackage[dvips]{graphicx}
\usepackage{subfigure}
\usepackage{bm}
\usepackage{caption}
\usepackage{stfloats}
\usepackage{epsfig}
\usepackage{algorithm}
\usepackage{algorithmic}
\usepackage[dvips]{color}
\usepackage{subeqnarray}
\usepackage{cases}
\usepackage{cite}

\begin{document}

\title{Joint Placement Optimization and RNC in UAV-based Wireless Multicast Networks
}

\author{\IEEEauthorblockN{Xianzhen Guo, Bin Li, Kebang Liu, Ruonan Zhang}\\
\IEEEauthorblockA{Department of Communication Engineering, Northwestern Polytechnical University, Xi'an, China 710072}
}

\maketitle

\begin{abstract}
Random network coding (RNC) is an efficient coding scheme to improve the performance of the broadband networks, especially for multimedia applications which are popular in 5G network. However, it is a challenging work to transmit the real-time media data because of the time limitation and wide band requirement. Moreover, the topology of the network changes due to users' movement, causing huge channel heterogeneity in large wireless network area. In this case, the fixed macro base station (BS) or access point may not fit the real-time user distributions. Accordingly, the UAV-based BS with high mobility can provide flexible service by adjusting it position according to users' locations to fit the dynamic topology of the network. Therefore, in this paper, we propose a UAV-based adaptive RNC (UARNC) scheme that jointly optimizes the UAV's location and RNC packet scheduling to maximize the throughput in a multicast network while guaranteeing the service quality of the bottleneck users. This problem is formulated as an optimization problem, and the greedy scheduling techniques and particle swarm optimization (PSO) algorithm are adopted to solve it. Finally, the simulation results prove the effectiveness of the proposed scheme.
\end{abstract}

\begin{IEEEkeywords}
Multicast network, multimedia, PSO, RNC, UAV
\end{IEEEkeywords}

\IEEEpeerreviewmaketitle

\begin{figure}[b]
	\vspace{-4mm}
	\footnotesize{This work is partially supported by National Natural Science Foundation of China under Grant 61571370, Key Research and Development Program of Shaanxi Province under Grant 2019ZDLGY07-10, Advance Research Program on Common Information System Technologies under Grants 315045204 and 315075702.
	}
\end{figure}

\IEEEpeerreviewmaketitle

\section{Introduction}

One of the factors that drive the development of the 5G network is the increasing demand for multimedia data transmission over broadband network \cite{mul}. Generally, the multimedia data should be transmitted to multiple users simultaneously. For spectrum efficiency, multicast technology can be integrated into the 5G ecosystem to satisfy the rapidly increasing demand for multimedia data and it will play an important role in emerging 5G networks \cite{mer}.

Some applications, e.g. mobile TV, live broadcast, real-time monitoring, are delay-sensitive and require high quality of service (QoS) to ensure smooth and timely video playback. To satisfy those requirements, a series of technologies, such as network coding (NC), can be adopted in multicast network. NC has been proved to provide a promising platform for the multicasting transmissions. In addition, the multicast capacity can be approached by using the random network coding techniques (RNC). In RNC, a set of the coding packets combined with the original data are transmitted to multiple users. However, in wireless mobile network, the locations of the users may change dynamically over time, causing the change of network topology and channel heterogeneity in the large network area. For example, in a certain period of the time, part of the users may form clusters around certain locations \cite{3d} which are far away from the base station (BS). In this case, the fixed terrestrial infrastructures can not provide high-quality services for them due to the long distance.

Fortunately, with high mobility, UAVs can move close to the users and provide better channel quality. Therefore, they have been widely applied in wireless communications \cite{survey1}, \cite{survey2}. In addition, some UAV-based multicast network have been proposed and studied \cite{add1,add2,add3,add4}. \cite{add1} characterizes the capacity of UAV-enabled multicast channel by jointly optimizing the UAV's trajectory and power allocation. A fly-and-communicate protocol is proposed in \cite{add2}, which designs the flying speed, UAV altitude and antenna beamwidth jointly for time minimization and energy minimization. However, few of them involves NC in the proposed multicast network. In \cite{add5}, the random linear network coding is adopted for transmission in UAV-enabled multicasting network while the authors only focus on the completion time minimization problem.

In this paper, we consider a UAV-based multicast network for multimedia data transmission. We propose an UAV-based adaptive RNC (UARNC) scheme which optimizes the UAV's location and RNC packet scheduling to maximize the network throughput. In UARNC, we use Markov Decision Process (MDP) to model the network dynamics and an optimization problem is formulated for the optimal location of the UAV-based base station (UBS) and the optimal packet scheduling for RNC. The optimization problem is solved by using two algorithms. A greedy scheduling techniques (GST) based algorithm is proposed to find the optimal actions in RNC scheme and another one is a PSO-based algorithm designed to optimize the UAV's location and RNC scheduling jointly. The simulation results show the effectiveness of the proposed methods.
Note that the boldface letters refer to vectors or matrices in this paper.

The rest of this paper is organized as follows. The system model is shown in Section~\ref{section:1}. The principle of ARNC is presented in Section~\ref{section:2}. Section~\ref{section:3} models the network dynamics as MDP and formulates the optimization problem. The proposed algorithms are introduced in Section~\ref{section:4}. Numerical results are shown in Section~\ref{section:5}. Finally, Section~\ref{section:6} concludes the paper.

\begin{figure*}[t]
\begin{center}
\epsfig{figure = 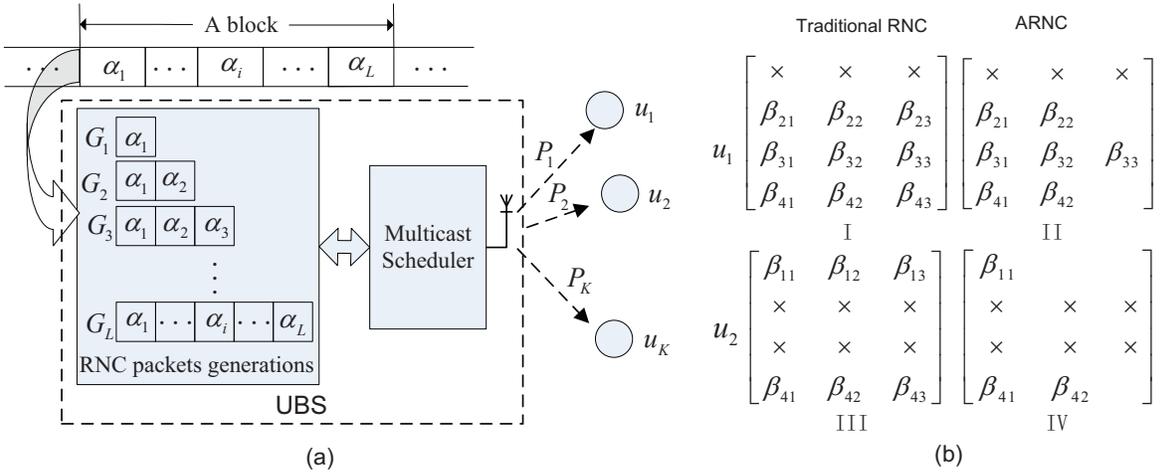, height=2.5in}
\caption{(a) System model; (b) users' encoding coefficient matrices. ``$\times$" means the network encoded packet has not successfully received by the user. Matrix I and II indicate packet loss at $t=1$ for user $u_1$ under two encoding methods. While matrixes III and IV show the packet loss at $t=2,3$ for user $u_2$.}
%\caption{a) System model.}
\label{fig:erasure_pattern}
\end{center}
\vspace{-0.1in}
\end{figure*}

\section{System Model}\label{section:1}

We consider a multicast network including $K$ users and one UBS. In order to provide multimedia streams with diversified QoS to users, scalable video coding (SVC) \cite{SVC} is used to partition the stream data into a base packet ($\alpha_1$) and several enhancement packets ($\alpha_2, \dots, \alpha_l, \dots, \alpha_L$). $\alpha_1$ is the most important one, which provides a reasonable video quality. The left packets are organized in a hierarchical fashion such that the $\alpha_{l-1}$ must be present for $\alpha_{l}$ to be useful. Under this structure, users can enjoy video with higher quality if they receive more enhancement packets. Naturally, the priority order of these multimedia data is $\alpha_1 \geq \alpha_2 \geq \dots\geq \alpha_l \dots\geq \alpha_L$, and the UBS should transmit $L$ prioritized packets to all users within $T$ time slots.

Here the UBS and all users are assumed to be with single antenna. In a three-dimensional (3D) Cartesian coordinate, the horizontal coordinate of user $i$ is denoted by $\boldsymbol{u_i}=(x_i, y_i)\in \mathbb{R}^{2\times1}$. The UBS is set to fly at a constant altitude $H$ and its horizontal coordinate is $\boldsymbol{q}=(x,y) \in \mathbb{R}^{2\times1}$. Therefore, the distance from the UBS to user $i$ is
\begin{equation}\label{eq1}
	 d_i = \sqrt{\Vert\boldsymbol{q}-\boldsymbol{u_i}\Vert_2^2+H^2}.
\end{equation}

%Assume $\boldsymbol{h_i}=(h_{i,1},\ldots,h_{i,N_t})\in \mathbb{C}^{N_t\times1}$ as the channel vector from the UBS to user $i$, $i=1,\ldots,K$ and $h_{i,m}$ as the channel coefficient from the $m$th antenna of UBS to user $i$.
According to \cite{zeng1}, the line-of-sight (LoS) model can provide a good approximation for the UAV-user channels. Thus, we assume that the links between the UAV and the users are dominated by the LoS channel. In addition, the Doppler effect caused by the UAV's continuous mobility is assumed to be perfectly compensated \cite{yongzeng}. As a result, the channel from the UBS to user $i$ is
\begin{equation}\label{eq2}
h_i=\frac{\beta_0}{d_i^2}=\frac{\beta_0}{\Vert\boldsymbol{q}-\boldsymbol{u_i}\Vert_2^2+H^2},
\end{equation}
where $\beta_0$ is the channel power when the distance from the UAV and user $i$ is 1m.

%The UBS with multiple antennas is assumed to form $M$ beams, resulting in $M$ multicast group. The transmit power allocate to group $m$ is $p_m$ with $\sum_{m=1}^{M}=P$, where $P$ is the total transmit power of UBS. Each user belongs to one and only one multicast group. Assume the multicast beamforming vector from the UBS to users in group $m$ as $\boldsymbol{w_m}\in\mathbb{C}^{N_t\times1}$ with $\Vert\boldsymbol{w_m}\Vert_2^2$$\leq p_m$. Denoted $\boldsymbol{W}=[\boldsymbol{w_1},\ldots,\boldsymbol{w_m}]$ as the beamforming matrix.
Then, the signal-to--noise ratio of user $i$ is expressed as:
\begin{equation}\label{eq3}
	\gamma_{i}=\frac{h_iP_t}{\sigma^2}=\frac{\beta_0P_t}{(\Vert\boldsymbol{q}-\boldsymbol{u_i}\Vert_2^2+H^2)\sigma^2},
\end{equation}
where $P_t$ is the transmitting power of the UBS. $\sigma^2$ denotes the white Gaussian noise variance at the users.

For simplicity, the encoded packets are assumed to be modulated by BPSK. Then, the bit error rate of user $i$ can be calculated as:
\begin{equation}\label{eq4}
P_i^b=Q(2\sqrt{\gamma_i}).
\end{equation}

The packet error rate (PER) $P_i^p$ of user $i$ is expressed as:
\begin{equation}\label{per}
P_i^p=1-(1-P_i^b)^{n},
\end{equation}
where $n$ is the length of the packet.

\newcounter{tmepEq}
\setcounter{tmepEq}{\value{equation}}
\setcounter{equation}{5}
\begin{figure*}[!b]
	\hrulefill
	\begin{equation}\label{pro_f}
f(l) = \left\{ \begin{array}{l}
{(1 - {\gamma _i})^l}{\gamma _i}\sum\limits_{i = 0}^{L - l - i} {C_{T - l - 1}^i{{(1 - {\gamma _i})}^i}{\gamma _i}^{T - l - 1 - i}, \quad 1 \le l < L}; \\
\sum\limits_{i = L}^T {C_T^i{{(1 - {\gamma _i})}^i}{\gamma _i}^{T - i},\quad \quad \quad \quad \quad \quad \quad \quad \quad \quad \quad \quad l = L}.
\end{array} \right.
	\end{equation}
\end{figure*}
At the beginning of each time slot, each user sends a one-bit feedback message via dedicated control channel to inform the BS on UAV whether the packet has been received successfully or not. However, due to its short length, we assume that the user feedback is error-free.

\section{Adaptive Random Network Coding}\label{section:2}

For easy understanding, we briefly introduce the principle of adaptive RNC (ARNC) proposed in our pervious work \cite{hongxiang}. For a SVC data block with $L$ prioritized packets, the UBS builds $L$ coding generators, in which different number of packets are encoded as a new coded packet. Specifically, each generator $G_l$ ($1\leq l \leq L$) consists of $l$ packets \{$\alpha_1, \alpha_2, \dots, \alpha_l$\} for encoding, as shown in Fig. \ref{fig:erasure_pattern}(a). At time slot $t$, the UBS linearly combines the packets using one of the $L$ generators. For example, when $L=3$, the network coded packets from the three generators are

\begin{enumerate}

\item[$\bullet$] $G_1$: combines only $\alpha_1$, i.e. $ c_{1,t} =\beta_{t1}\alpha_1$;
\item[$\bullet$] $G_2$: combines $\alpha_1$ and $\alpha_2$, i.e. $c_{2,t}=\beta_{t1} \alpha_1+ \beta_{t2} \alpha_2$;
\item[$\bullet$] $G_3$: combines $\alpha_1$, $\alpha_2$ and $\alpha_3$, i.e. $c_{3,t}=\beta_{t1} \alpha_1+\beta_{t2} \alpha_2 +\beta_{t3} \alpha_3$,
\end{enumerate}
where $\beta_{ti}$ (i=1,2,3) are the encoding coefficients randomly drawn from a large finite field $\mathbb{F}_q$.

Unlike the traditional RNC in which the coding packet combines all original packets, in ARNC the number of packets for encoding is variable and the user needs only partial set of the coding packets to decode out useful information. To explore the advantage, we take the example that each data block contains $L=3$ packets ($\alpha_1 \geq \alpha_2 \geq \alpha_3$) to be delivered to two users $u_1$ and $u_2$ within $T=4$ time slots under the erasure channel. Fig. \ref{fig:erasure_pattern}(b) shows the RNC encoding coefficient matrices received by different users.

Under the scheme of traditional RNC, even though $u_1$ losses the first coding packets, its coefficient matrix is still full rank to decode out all packets. While for $u_2$, it just receives two coding packets in the 1st and 4th slots so that it cannot decode out any packet. Therefore, the network throughput is $3/4$ packet per time slot.

In ARNC, Fig. \ref{fig:erasure_pattern}(b) demonstrates that $u_1$ collects a full set of the encoded packets to decode out all the packets. It also shows that $u_2$ can still decode $\alpha_1$ and $\alpha_2$ with only a partial set of the coded packets using ARNC scheme. As a result, this scheme achieves a network throughput of $5/8$ per time slot, which is $67\%$ higher than that of the traditional RNC.

Without loss of the generality, when packet in one block is $L$, the probability that each user can decode out the first $l$ packets is shown in \eqref{pro_f} according to \cite{hongxiang}.

\section{Problem Formulation}\label{section:3}

In this paper, we expect to optimize the location of UBS so as to maximize the throughput while considering ARNC scheme in the UAV-based wireless multicast network.
Given a 2-D location of UBS $\boldsymbol{q}$ and its flight height $H$. At the beginning of each time slot, the UBS transmits the first packet from the generator $G_1$. Because of the lossy channel with certain PER, the network status changes in each time slot. Therefore, the UBS needs to encode the packets using ARNC for scheduling to maximize the network throughput based on real-time network status. Clearly, for a certain UBS location, the network status in the next time slot depends on the current network state and the scheduling action during the current time slot. Accordingly, the network dynamics can be modeled as a MDP, in which the UBS decides the optimal action to take at each time slot. Specifically, we use parameters $(\bm{q},\bm{A}, \bm{S}, r, T)$ to specify the network dynamics.

\begin{enumerate}
	
\item UAV's location $\bm{q}$: for a given location $\bm{q}$, the distance between the UAV and each user can be obtained according to \eqref{eq1}. Substituting \eqref{eq1}, \eqref{eq2}, \eqref{eq3} and \eqref{eq4} into \eqref{per}, the PER of each user can be obtained.
	
\item Scheduling mode set $\bm{A}$: it includes a set of scheduling modes to multicast the coding packets from different generators. In detail, the $G_l$ ($1\leq l \leq L$) consists of $l$ packets \{$\alpha_1, \alpha_2, \dots, \alpha_l$\} for encoding. The coding packet $g_l$ created by $G_l$ is a linear combination of the packets in one block.

\item Status matrix $\mathbf{s}_{i,t}$: $\mathbf{s}_{i,t}$ is a $L\times T$ status matrix that denotes the status of the coding packets received by the user $i$. The $j$th row of $\mathbf{s}_i$ contains the coding coefficient vector received at $j$th time slot. For example: when we set $L=3$ and $T=5$, $\mathbf{s}_{i,t}$ may be shown as
\setcounter{equation}{6}
\begin{equation}\label{EQ.2}
\begin{split}
\bm{s}_{i,t}= \left[ \begin{array}{ccccc} \beta_{11} & \beta_{12}& 0&  \beta_{14} & 0 \\
                                              0       & \beta_{22}& 0&  \beta_{24} & 0  \\
                                              0       &0           & 0&  \beta_{34} & 0  \\
  \end{array} \right].
\end{split}
\end{equation}

It indicates that user $i$ successfully receives three coding packets ($c_{11}$, $c_{22}$, $c_{34}$) from $G_1$ at $t=1$, $G_2$ at $t=2$, and $G_3$ at $t=4$, respectively. Obviously, it can successfully decode out $\alpha_1$, $\alpha_2$ and $\alpha_3$ from $\bm{s}_{i,t}$.

\item Network state matrix set $\bm{S}$: it shows the packet receiving status set of the overall network. $\bm{S}_t$ ($\bm{S}_t\in \bm{S}$) gives the network status at time slot $t$, which is defined as $\bm{S}_t=\mathop \cup \limits_{i = 1}^{\rm{K}}\bm{s}_{i,t}$.

\item Immediate reward $ r(\bm{S}_{t},\bm{q},a_t)$: it represents the reward (i.e., the network throughput improvement) by taking action $a^t(a^t \in \bm{A})$ for the given $\bm{q}$ and $\bm{S}_{t}(\bm{s}^{t} \in  \bm{S})$ at $t$. This reward can be calculated by
	\begin{equation}\label{reward}
	\begin{split}
	r(\bm{S}_{t},\bm{q},a^t)&=\mathbb{E}\left[ r(\bm{S}_{t+1}|\bm{S}_{t},\bm{q},a_t)\right]\\&= \sum_{i=1}^{K} \mathbb{E}\left[ r(\bm{s}_{i,t+1}|\bm{s}_{i,t},\bm{q},a_t)\right],
	\end{split}
	\end{equation}
where $\mathbb{E}[\cdot]$ denotes the expectation with respect to $\bm{S}_{t+1}$; $\bm{q}$ is the given location of UAV; vector $\bm{s}_{i,t}$ and $\bm{s}_{i,t+1}$ indicate the network status of $u_i$ at $t$ and $t+1$; $r(\bm{s}_{i,t+1}|\bm{s}_{i,t},\bm{q},a_t)$ is the function that calculates the future-dependent reward $u_i$ from $\bm{s}_{i,t}$ to $\bm{s}_{i,t+1}$ for the given $\bm{q}$ and $a_t$. Note that under a given $\bm{s}_{i,t}$, $\bm{s}_{i,t+1}$ only has two states: receiving the encoded packet (denoted as $\bm{s}_{i,t+1}(1)$) or not (denoted as $\bm{s}_{i,t+1}(0)$). For instance, if the state of $u_i$ is
$$\bm{s}_{i,t}= \left[ \begin{array}{ccccc} \beta_{11} & \beta_{12}& 0&  0 & 0 \\
                                              0       & \beta_{22}& 0&  0& 0  \\
                                              0       &0           & 0&  0 & 0  \\
  \end{array} \right]$$
at $t$, and the BS takes scheduling action $a_t=$ {``sending a coding packet $g_3$ from $G_3$''}, the probability for the next state being
$$ \bm{s}_{i,t}= \left[ \begin{array}{ccccc} \beta_{11} & \beta_{12}& \beta_{13}&  0 & 0 \\
                                              0       & \beta_{22}& \beta_{23}&  0& 0  \\
                                              0       &0           & \beta_{33}&  0 & 0  \\
  \end{array} \right]$$
is expressed as
$$p(\bm{s}_{i,t+1}(1)|\bm{s}_{i,t},\bm{q},a_t)=1-p_{i}^{p}(\bm{q})$$ and the immediate reward is $r(\bm{s}_{i,t+1}(1)|\bm{s}_{i,t},\bm{q},a_t) = 3-2=1$.
Therefore, \eqref{reward} can be rewritten as
\begin{equation}\label{111}
\begin{split}
&r(\bm{S}_{t},\bm{q},a_t)=\\&
\sum_{i=1}^{U} \left(\sum_{i=0}^{1} \left(1-p_{i}^{p}(\bm{q})\right)^{i}\left(p_{i}^{p}(\bm{q})\right)^{(1-i)}r(\bm{s}_{i,t+1}(i)|\bm{s}_{i,t},a_t)
	\right).
\end{split}
\end{equation}

\item Deadline $T$ ($T\geq L$): it is the number of time slots available for transmitting the packet block.
\end{enumerate}

Based on the above MDP framework, a transmission policy $\bm{\Omega}$ is well scheduled by optimizing $a_t \in \bm{A}$ and UBS's location $\bm{q}$.
Let $\Gamma_{\bm{\Omega}}$ be the expected reward obtained by following policy $\bm{\Omega}$, which is defined as
	\begin{equation}\label{fit2}
	\Gamma_{\bm{\Omega}} (\bm{S}_0) = \frac{1}{T}\sum_{t=0}^{T-1}r(\bm{s}_{i,t+1}|\bm{s}_{i,t},\bm{q},a_t),
	\end{equation}
where $\bm{\Omega} =(a_0,\dots, a_t,\dots, a_{T-1}, \bm{q})$ and $\bm{S}_{0}$ is the initial network state.

Note that $\Gamma_{\bm{\Omega}}$ is the average network throughput per time slot. The goal of the transmission policy is to find an optimal $\bm{\Omega}^*=[(a_0)^*,\dots, (a_t)^*,\dots, (a_{T-1})^*, \bm{q}^*]$ during the $T$ steps that can maximize $\Gamma_{\bm{\Omega}}$, where $\bm{q}^*$ is the optimal location of UAV and $(a_t)^*, t=0,1,\dots, T-1$ is the optimal action at each time slot.

In addition, for network fairness, we should guarantee the service quality of the bottleneck users which have the worst channel qualities. According to the characteristics of the SVC data, each user can get the basic video quality if they can receive the base packet. The more enhancement packets they get, the higher quality of the multimedia data they enjoy.
Therefore, we assume that each user should receive at least the first $l$ packets with higher priority to maintain the basic QoS.
According to \eqref{pro_f}, the probability that each user can decode out at least the first $l$ packets is expressed as:
\begin{equation}
	P_{i,l}=\sum_{i=l}^{L}f(l).
\end{equation}

Then, we set a value $P_{th}$ as the fairness transmission threshold. Clearly, $P_{i,l}$ should be no less than $P_{th}$ and the optimization problem can be formulated as:
\begin{equation}\label{opti}
\begin{split}
(h^*,r^*)=&\underset{h,r}{\arg \max}\sum_{1}^{K}C_i,
\\& \text{s. t. }
P_{i,l}\geq P_{th},\quad i=1,\dots,K.
\end{split}
\end{equation}

\section{Optimization Algorithms}\label{section:4}
Equation \eqref{opti} is a joint optimization problem with multiple constraints and variables. It is difficult to get the optimal results directly. To simplify this problem, we divide the solution of this optimization problem into two steps.
Firstly, we adopt a low-complexity greedy scheduling technique (GST) \cite{hongxiang} in Algorithm~\ref{rew} to find the optimal $(a_t)^*, t=0,1,\dots, T-1$ for a given position of UAV $\bm{q}_0$.
In GST, based on the network status from the users, the UBS finds the appropriate $a_t$ in each time slot to maximize the reward $  r(\bm{S}_{t},\bm{q}_0,a_t)$. The output of the algorithm is the optimal action for each time slot and the average reward $\Gamma_{\bm{\Omega}}$.
\begin{algorithm}[h]
	\caption{: Algorithm for the optimal actions}
	\label{rew}
	{{\bf Input}: $ \bm{A}, \bm{q}_0, T, L, p_i$. \\
	{\bf Output}: $(a_t)^*, t=0,1,\dots, T-1$ and $\Gamma_\Omega$}

	\begin{algorithmic}[1]
		\STATE{Initialize: $\bm{s}_{0} = [\bm{0}]$ and $r(\bm{s}_{0})  =0$ }
		\FOR{$t = 0$ to $T-1$}
		\STATE{$(\bm{a}_{t})^*= \arg \max\limits_{\bm{a}_{t} \in \bm{A}} r(\bm{S}_{t},\bm{q}_0,a_t) $}
		\ENDFOR
		\STATE{$\Omega_0=[(a_0)^*,\dots, (a_t)^*,\dots, (a_{T-1})^*, \bm{q}_0]$}
		\STATE{$\Gamma_{\bm{\Omega}}=\Gamma_{\bm{\Omega}_0}$}
	\end{algorithmic}
\end{algorithm}

According to Algorithm~\ref{rew}, we can get the optimal actions and the average reward with a given location of UBS.

Next, we adopt the particle swarm optimization algorithm (PSO) \cite{PSO} to find the optimal location of UBS.
The PSO algorithm consists of a population of particles that move over a search space. Each particle is a candidate solution for the optimization problem. The position and velocity of each particle is denoted by $\boldsymbol{o_i}$ and $\boldsymbol{v_i}$ respectively.
During the searching process, each particle will update its position and velocity according to \eqref{upv} and \eqref{upp} respectively to find the best solution. In our paper, the particle's position is limited by the constraints in \eqref{opti}. So before updating the particle's position, we need to check if the new position satisfies the constraints, i.e. $P_{i,l}\geq P_{th}, i=1,\dots,K$.

In order to find the global best, we assume that each particle will share their local best with other particles.
\begin{equation}\label{upv}
\begin{split}
\boldsymbol{v_i}(t+1)=&w\boldsymbol{v_i}(t)+C_1\phi_1(\boldsymbol{P_i}(t)-\boldsymbol{o_i}(t))\\&+C_2\phi_2(\boldsymbol{P_{gi}}(t)-\boldsymbol{o_i}(t)),
\end{split}
\end{equation}
\begin{equation}\label{upp}
\boldsymbol{o_i}(t+1)=\boldsymbol{o_i}(t)+\boldsymbol{v_i}(t+1),
\end{equation}
where $\boldsymbol{v_i}(t)$ and $\boldsymbol{o_i}(t)$ are the velocity and current position of the particle $i$; $w$ is a random parameter; $C_1$ and $C_2$ represent the intensity of the attraction of a particle towards its local best and global best respectively; $\boldsymbol{P_i}(t)$ and $\boldsymbol{P_{gi}}(t)$ denote the local best and the global best of a particle respectively; $w$ is the inertial coefficient of each particle.

Generally, a fitness function is defined in PSO. We aim to find the optimal optimization for maximum network throughput. In this paper, the fitness is the maximal average reward and it can be obtained by Algorithm~\ref{rew}. The detailed process of this PSO algorithm is shown in Algorithm~\ref{alg}.
\begin{algorithm}
	\renewcommand{\algorithmicrequire}{\textbf{Input:}}
	\renewcommand{\algorithmicensure}{\textbf{Onput:}}
	\caption{PSO algorithm description for the best position}
	\label{alg}	
	\begin{algorithmic}	
		\REQUIRE $w$, $C_1$, $C_2$, $\phi_1$, $\phi_2$, the number of time slot $T$ and the packet number $L$
		\ENSURE The optimal position of UAV
		\FOR{each particle $i$}
		\STATE Initialize velocity $\boldsymbol{v}_i$ and the position $\boldsymbol{o}_i$ for particle $i$
		\STATE Evaluate particle $i$ and set $\boldsymbol{P}_i=\boldsymbol{o}_i$
		\ENDFOR
	    \STATE find the global best $\boldsymbol{P}_{gi}$
		\FOR{j=1;~j\textless maxg;~j++}
		\FOR{n=1;~n\textless sizepop;~n++}
		\STATE Update the velocity and position of particle $i$
		\STATE Evaluate particle $i$
		\STATE Find the optimal action for each particle $i$ using Algorithm~\ref{rew}
		\STATE Calculate the fitness of each particle according to \eqref{fit2}
		\IF{fit($\boldsymbol{o}_i$) \textgreater fit($\boldsymbol{P}_i$)}
		\STATE $\boldsymbol{P}_i$=$\boldsymbol{o}_i$
		\ENDIF
		\IF{fit($\boldsymbol{o}_i$) \textgreater -fit($\boldsymbol{P}_{gi}$)}
		\STATE $\boldsymbol{P}_{gi}$=$\boldsymbol{o}_i$
		\ENDIF	
		\ENDFOR
		\ENDFOR
		\STATE print $\boldsymbol{P}_{gi}$	
	\end{algorithmic}
\end{algorithm}

In Algorithm~\ref{alg}, `fit($\boldsymbol{o}_i$)' and `fit($\boldsymbol{P}_i$)' denote the fitness of $\boldsymbol{o}_i$ and $\boldsymbol{P}_i$. `sizepop' and `maxg' are the number of the particles and literation.

\section{Numerical Results}\label{section:5}
 Assume all the users distribute in a 1000m$\times$1000m area and the UAV's flight altitude is fixed at 200m.
%To measure the heterogeneity of the user distribution, we adopt the Coefficient of Variation (CoV) of the Voronio cell area proposed in \cite{3d}. The CoV is defined as $C_v$. When $C_v=1$, the users are distributed in the area uniformly. Otherwise, the users will accumulate into clusters around some hotpots.
We assume that $C_1$=$C_2$=1.4955, maxg=400, sizepop=100, P=25 mW, $\beta_0$=-70dB, $\delta$=-150dB and the packet length $n$=10.
To illustrate the effectiveness of our scheme, we compare the throughput performance of UARNC with those of other schemes including Random Network Coding (RNC), Automatic Repeat Request (ARQ) and Round-Robin Scheduling (RRS).

\begin{figure}[htbp]
	\centering
	\subfigure[The optimal position of UBS.]{\includegraphics[width=4.55cm]{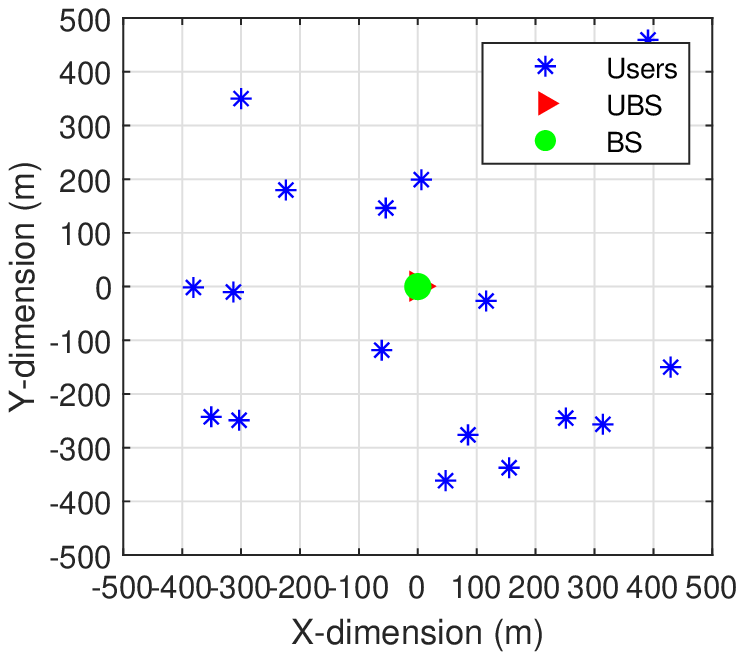}\label{1}}
\hspace{-5mm}
	\centering
	\subfigure[Iteration process.]{\includegraphics[width=4.55cm]{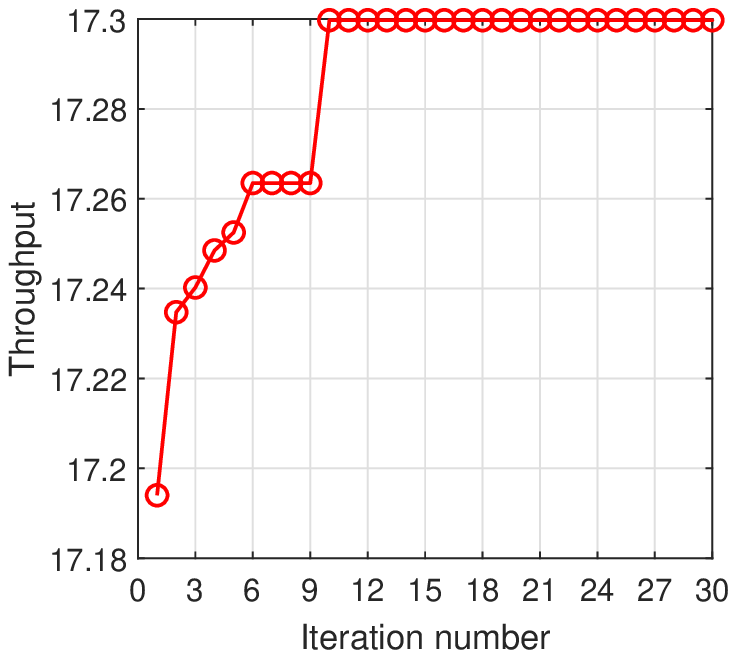}\label{2}}
	\caption{The optimal position of UBS and the algorithm iteration process with uniformly distributed users.}
	%\label{fig1}
\end{figure}

\begin{figure}[htbp]
	\centering\subfigure[The optimal position of UBS.]{\includegraphics[width=4.55cm]{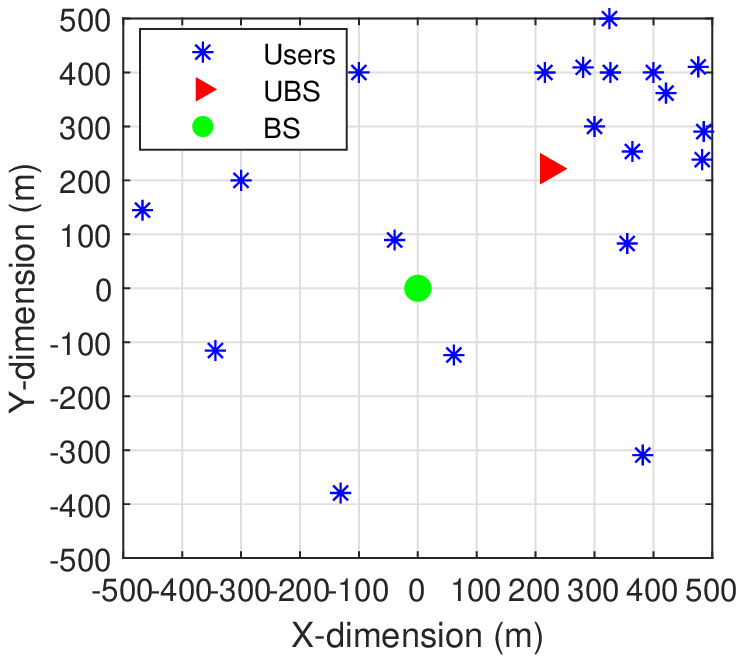}\label{3}}
\hspace{-5mm}
	\centering\subfigure[Iteration process. ]{\includegraphics[width=4.55cm]{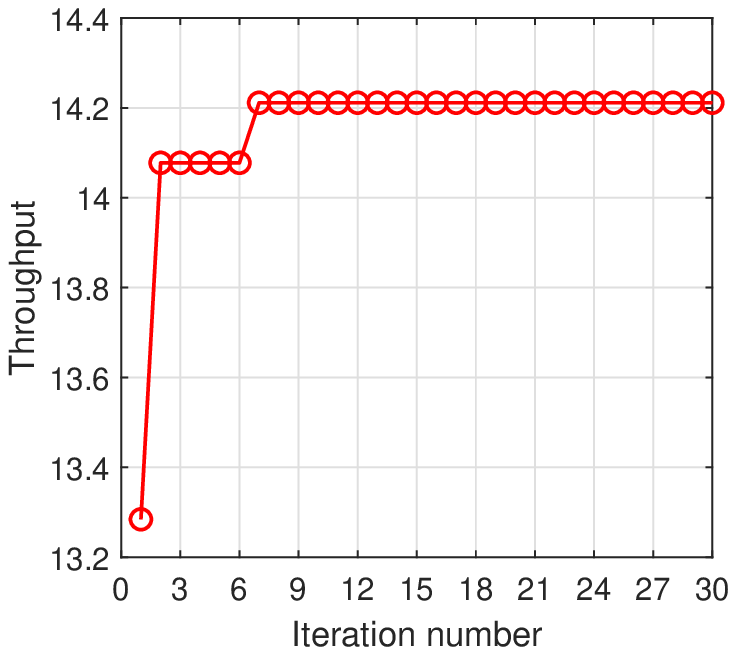}\label{4}}
	\caption{The optimal position of UBS and the algorithm iteration process when users cluster around some hot spots.}
	%\label{fig2}
\end{figure}
%\begin{figure}
%	\centering
%	\includegraphics[width=10cm]{dis1.eps}
%	\caption{The optimal position of UBS and the algorithm iteration process with uniformly distributed users.}
%	\label{fig1}
%\end{figure}

Fig.~\ref{1} presents the optimal location of the UAV when users distribute uniformly in the area. The blue asterisks, red triangle and green filled circle represent the locations of users, the UBS and a fixed BS respectively. A fixed BS is assumed to be at the center of the area.
In this case,  we can see that the optimal location of UBS is close to the center of the area.

Fig.~\ref{3} shows the corresponding results when the users accumulate around some hotpots. Comparing with the results in Fig.~\ref{1}, we see that the UBS adjusts its location to somewhere near to the hotpots for maximum network throughput. However, the BS, fixed at (0,0), cannot adjust its location according to the change of network topology.

%\begin{figure}
%	\centering
%	\includegraphics[width=10cm]{dis2.eps}
%	\caption{The optimal position of UBS and the algorithm iteration process when users cluster around some hot spots.}
%	\label{fig2}
%\end{figure}
In addition, the iteration process of the proposed algorithms in those two scenarios are demonstrated in Fig.~\ref{2} and Fig.~\ref{4}. From the results, we can see that the algorithms can solve the problem effectively. The optimal location can be found within 10 and 7 times iteration with different user distributions respectively.

\begin{figure}
	\centering
	\includegraphics[width=8cm]{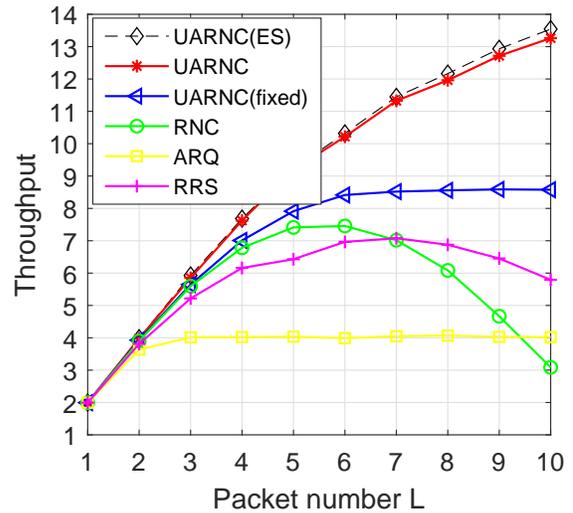}
	\caption{Network throughput vs. packet number L.}
	\label{fig3}
\end{figure}

In Fig.~\ref{fig3}, the average network throughput in term of $L$ with $T=10$ is presented.
Firstly, we assume that the UBS is hovering over a fixed location (0,0) which means that we only adopt the ARNC without optimizing the UAV's location. The yellow line with a legend 'UARNC(fixed)' shows the corresponding results. In addition, the performance of other schemes, i.e. RNC, ARQ, RRS, are also plotted in this figure. Comparing with the results of those schemes, our proposed scheme performs better even without optimizing the location of UBS.
Then, we give the results (red line) of UARNC with jointly optimizing UAV's location and ARNC scheduling. Comparing the results with others, we can see that the proposed UARNC outperform all other schemes, especially with multiple packets.

Moreover, the dark dashed line with a legend 'UARNC(ES)' represents the results obtained using Exhaustive Search (ES) algorithm.
The performance gap of the red line and the dark line is negligible, which proves the effectiveness of the proposed algorithm.

\begin{figure}
	\centering
	\includegraphics[width=8cm]{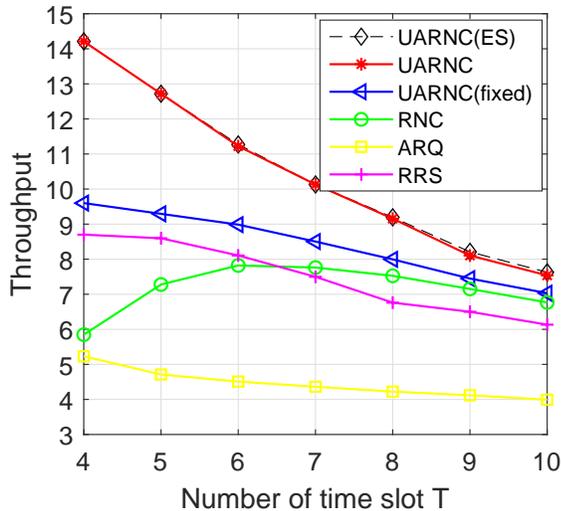}
	\caption{Network throughput vs. number of time slot T.}
	\label{fig4}
\end{figure}

Fig. \ref{fig4} plots the network throughput of different schemes as a function of $T$ with $L=4$.
Similarly, we give the results of four schemes (UARNC, RNC, ARQ, RRS).
The average network throughput of three schemes (UARNC, ARQ, RRS) declines with the increasing of $T$.
Generally, users can decode more packets with more time slots. However, the packet number is fixed at 4 in this scenario. Thus, there must be a optimal $T$ at which the network throughput is the biggest. Beyond the optimal $T$, the average throughput will decrease and it (averaged over $T$) converges to zero when $T$ approaches infinity. For those three schemes, the optimal $T$ is 4 when $L=4$. However, the optimal $T$ is 6 for RNC. Therefore, the throughput of this scheme first increases to the biggest and then decreases with increasing number of time slot.

\section{Conclusions}\label{section:6}

In this paper, we consider a RNC multicast network where there are multiple users and one UAV-based BS that fully explores the high mobility to provide flexible service to the users. To fit this requirement, an UARNC scheme that jointly optimizes the UAV's location and RNC packet scheduling to maximize the throughput in multicast network is proposed. Then, two algorithms based on greedy scheduling technique and PSO respectively are proposed to solve the optimization problem. The proposed UARNC outperforms other schemes, such as RNC, ARQ and RRS.

\bibliographystyle{ieeetran}
\bibliography{arnc}
\end{document}